\documentclass[12pt,preprint]{aastex}

\shortauthors{Middleditch}
\shorttitle{87A Link to SNe \& GRBs}

\begin{document}

\title{The SN 1987A Link to Others and Gamma-Ray Bursts
%\newline \today\ at \the\time\ minutes VERSION
}

\author{John Middleditch$^1$}

\altaffiltext{1}{Modeling, Algorithms, \& Informatics, CCS-3, MS B265,
                 Computer, Computational, and Statistical Sciences 
		 Division, Los Alamos National Laboratory, Los Alamos, 
		 NM 87545; jon@lanl.gov}

\email{jon@lanl.gov}

\begin{abstract}

Early measurements of SN 1987A can be interpreted in light of the
beam/jet (BJ), with a collimation factor $>$10$^4$, which had to 
hit polar ejecta (PE) to produce the ``Mystery Spot'' (MS), some 
24 light-days distant.  Other details of SN 1987A strongly suggest 
that it resulted from a merger of two stellar cores of a common 
envelope (CE) binary, i.e. a ``double degenerate'' (DD)-initiated 
SN.  Without having to blast through the CE of Sk -69$^{\circ}$ 202, 
it is likely that the BJ would have caused a full, long-soft gamma-ray 
burst ($\ell$GRB) upon hitting the PE, thus DD
can produce $\ell$GRBs.  Because DD must be the overwhelmingly 
dominant merger/SN mechanism in elliptical galaxies, where only short, 
hard GRBs (sGRBs) have been observed, DD without CE or PE must also 
produce sGRBs, and thus the pre-CE/PE impact photon spectrum of 99\% 
of {\it all} GRBs is {\it known}, and neutron star (NS)-NS mergers may 
not make GRBs as we know them, and/or be as common as previously 
thought.  Millisecond pulsars (MSPs) in the non-core-collapsed 
globular clusters are also 99\% DD-formed from white dwarf (WD)-WD 
merger, consistent with their 2.10 ms minimum spin period, the 2.14 ms 
signal seen from SN 1987A, and sGRBs offset from the centers of 
elliptical galaxies.  The many details of Ia's strongly suggest that 
these are also DD initiated, and the single degenerate total 
thermonuclear disruption paradigm is now in serious doubt as well.  
This is a cause for concern in Ia Cosmology, because Type Ia SNe will 
appear to be Ic's when viewed from their DD merger poles, given 
sufficient matter above that lost to core-collapse.  As a DD-initiated 
SN, 1987A appears to be the Rosetta Stone for 99\% of SNe, GRBs and MSPs, 
including all recent nearby SNe except SN 1986J, and the more distant SN 
2006gy.  There is no need to invent exotica, such as ``collapsars,'' to 
account for GRBs.

\end{abstract}


\section{Introduction}

In Supernova 1987A (87A), Nature has provided an unparalleled
opportunity to learn the details of one of the most frequent, and 
violent events in the Universe.  Although confirming some early 
expectations of theorists \citep{Chev92b}, even from the first, 
features which would never have been seen at ordinary extra-galactic 
distances, appeared in the early light curve, which at that time 
defied easy explanation.

The most remarkable feature\footnote{Not counting, for the moment, 
the 2.14 ms pulsed optical remnant, which also revealed a $\sim$1,000 
s precession (Middleditch et al.~2000a,b -- hereafter M00a,b).  Since 
a prototypical, dim, thermal neutron star remnant (DTN) has been 
discovered in Cas A \citep{Ta99}, representing what PSR 1987A will look 
like after another 300 years, and other pulsars have since been observed 
to precess \citep{St00}, this candidate is no longer controversial.}
of 87A was the ``Mystery Spot'' (MS), 
with a thermal energy of 10$^{49}$ ergs, even 50 days {\it after} 
the core-collapse (CC) event (Meikle et al.~1987; Nisenson et al.~1987)
and separated from the SN photosphere ``proper'' (PP) by some 0.06 arc s,
with about 3\% of this energy eventually radiated
in the optical band.  
The possibility that this enormous energy implied for the MS might somehow 
link it gamma-ray bursts (GRBs) generally went unnoticed at the time.

GRBs, particularly long, soft GRBs ($\ell$GRBs), appear to be the most 
luminous objects in the Universe, occurring at the SN rate of one per 
second, given a collimation factor near 10$^5$, yet we still know very 
little about them (see, e.g., M\'esz\'aros 2006 and references therein).  
Although some have been found to be associated with SNe, others, mostly 
those lasting only a fraction of a second, with slightly harder spectra 
(sGRBs), produce only ``afterglows,'' sometimes extending down to radio 
wavelengths.  A large number of models have been put forth to explain 
GRBs, including NS-NS mergers for sGRBs, and exotic objects such as 
``collapsars'' \citep{MW99} for $\ell$GRBs.  The prime physical 
motivation for these is the enormous energy of up to 10$^{54}$ ergs 
implied for an isotropic source.  However, given that the data
from 87A presented herein support a beam/jet (BJ) collimation 
factor (CF) $>$10$^4$ in producing its MS (see $\S$\ref{sec:early}), 
there is no need for such a high energy.

This letter offers a simple explanation for 99\% of SNe, MSPs, and 
GRBs,\footnote{All except Soft Gamma Repeater [SGR] GRBs, which are 
estimated to amount to less than 5\% of sGRBs and 1.5\% of the total 
\citep{Pa05}.} in the context of 
the DD SN 1987A, its BJ and MS (Middleditch 2004, hereafter 
M04).  It further argues that these start as sGRBs, and only 
later are some modified to $\ell$GRBs (and one other type -- see 
$\S$\ref{sec:link}), by interaction with the common 
envelope (CE) and/or polar ejecta (PE).  
It also argues that many, possibly all SNe Ia are caused by DD 
(merger-induced) CC, the single degenerate (SD) paradigm (total 
thermonuclear disruption) being now admittedly in serious doubt 
(Siegfried 2007).  
Thus Ia Cosmology has not yet successfully challenged the Standard 
Model, and the burden of proof, for an accelerating expansion of the 
Universe, lies with the challenging model, the convenience of Concordance
Cosmology amounting to only that.  

\section{The SN 1987A Bipolarity and ``Mystery Spot''}
\label{sec:bipolar}

SN 1987A is clearly bipolar (NASA et al.~2007; 
Wang et al.~2003).  A ``polar blowout feature'' (PBF -- a needed
candidate for the r-process, e.g., Arnould et al.~2007) approaches 
at $\sim$45$^{\circ}$ off our line of sight, partially 
obscuring an equatorial bulge/ball (EB), behind which 
part of the opposite, receding PBF is visible.  The 87A PBFs 
and EB are approximately equally bright, in contrast to what
polarization observations imply for Type Ia SNe (see 
$\S$\ref{sec:Ia/c}).

A binary merger of two electron degenerate stellar cores 
(DD -- in isolation these would be white dwarfs [WDs]) has been 
proposed for 87A \citep{Pod89}, and the triple ring structure has 
recently been calculated in this framework \citep{MP07}.
Many other details of 87A, including the mixing (Fransson et 
al.~1989), the blue supergiant progenitor, the early polarization 
(Schwarz \& Mundt 1987; Barrett 1988), and the 2.14 ms optical 
pulsations (M00a,b), strongly support this hypothesis.  

The first clear evidence for DD-formed MSPs coincidentally came in the 
birth year of 87A, with the discovery of the 3 ms pulsar, B1821-24 
\citep{Ly87}, in the non-core-collapsed (nCCd) globular cluster (GC) M28.
Subsequently many more were found in the nCCd GCs, such as
47 Tuc, over the next 20 years, and attributing these to recycling through 
X-ray binaries has never really worked \citep{CMR93}, by a few orders of 
magnitude.\footnote{Relatively slowly rotating, recycled pulsars weighing 
1.7 M$_\odot$, in the CCd GC, Ter 5 \citep{Ra05}, have removed high 
accretion rate from contention as a alternative mechanism to produce the 
MSPs in the nCCd GCs.  The three MSPs in Ter 5 with periods $<$ 2 ms, Ter 
5 O, P, and ad \citep{He06}, may have been recycled starting with periods
near 2 ms.  There are four in this sample with periods between 2.05 and 
2.24 ms, and another, the first from Arecibo ALFA, at 2.15 ms \citep{Ch07}.}  

The 0.059 arc s offset of the MS from the
PP {\it coincides} with the PBF bearing of 194$^{\circ}$ (and thus
{\it along the axis of its DD merger}),\footnote{The far-side (southern)
minor axis of the equatorial ring has a bearing of 179$^{\circ}$.} 
some 45$^{\circ}$ off our line of sight, corresponding to 24 light-days 
($\ell$t-d), or 17 $\ell$t-d in projection, it taking light from 87A 
only {\it eight} extra days to reach the Earth after hitting the MS,
and there is 
evidence for exactly this delay (see below).
In addition, 
the typical 0.5$^{\circ}$ collimation for an
$\ell$GRB, over the 24 $\ell$t-d from 87A to its PE, produces $\sim$100 s 
of delay, within the range of the non-prompt components of $\ell$GRBs.  

\section{The Early Luminosity History of SN 1987A}
\label{sec:early}

The early luminosity histories of 87A 
taken with the Cerro Tololo Inter-American Observatory (CTIO) 
0.41-m \citep{HS90} and the Fine Error Sensor (FES) of the 
International Ultraviolet Explorer (IUE) \citep{Wa87}, both show 
such evidence of the BJ and MS (Fig.~1).\footnote{The CTIO V band 
center occurs at 5,500 \AA, as opposed to 5,100 
\AA ~for the FES, and in consequence, the FES magnitudes have 
been diminished by 0.075 in Fig.~1 to account for the resulting 
luminosity offset.}

Following the drop from the initial flash, the luminosity rises 
again to a maximum (`A' in Fig.~1) of magnitude 4.35 at day
3.0, interpretable as the hotter, more central part of the 
BJ shining through/running ahead of the cooler, roughly 
cylindrical outer layers which initially shrouded it.  This declines 
(`B') to magnitude 4.48 around day 7.0, interpretable as 
free-free cooling of, or the loss of the ability to cool for, an 
optically thin BJ.
The initial flash 
should scatter in the PE at day 8, and indeed `C' shows 
$\sim$2$\times$10$^{39}$ ergs/s in the optical for a day at day 8.0, 
and a decline {\it consistent with the flash} after that, indicating
a CF $>$10$^4$ for this component (beam).  
A linear ramp in luminosity starting 
near day 10 indicates particles from the BJ penetrating into the PE, 
with the fastest traveling at $>$0.9 c, {\it and} a particle CF $>$10$^4$.  
A decrement\footnote{This is 
preceded by a spike of up to 10$^{40}$ ergs s$^{-1}$ in the CTIO data, 
with the unusual colors of B, R, and I, in ascending 
order.  Optical pulsations were not seen during 
this early period (R.~N.~Manchester, private communication, 2007).
The possibility of less-than-coherent pulsations, though, is harder 
to eliminate.}  
of $\sim$5$\times$10$^{39}$ ergs s$^{-1}$ appears in both 
data sets near day 20 (`D').
The CTIO point just before the decrement
can be used as a rough upper limit for the MS luminosity, and 
corresponds to an 
excess above the minimum (near day 7.0) of 5$\times$10$^{40}$ ergs 
s$^{-1}$, or magnitude 5.8, the {\it same} as that observed in H$\alpha$ 
for the MS at days 30, 38, and 50.
%about 23\% of the total optical flux of 2.1$\times$10$^{41}$ ergs s$^{-1}$
%at that time.

\section{The SN 1987A link to GRBs}
\label{sec:link}

Without the H and He in the envelope of the progenitor of 87A, 
Sk -69$^{\circ}$202, the collision of the BJ with its PE (which 
produced the MS) might be indistinguishable from a full $\ell$GRB
\citep{Cen99}.\footnote{Otherwise it would just beg the question
of what distant, on-axis such objects would look like.}
This realization, together with the observation that 
no $\ell$GRBs have been found in elliptical galaxies, and 
the realization that the DD process {\it must} dominate (as 
always, through binary-binary collisions) by a large factor the NS-NS 
mergers in these populations, even when requiring enough WD-WD merged 
mass to produce CC, leads to the {\it inescapable} conclusion that 
the DD process produces sGRBs in the absence of CE and PE, the means 
by which they would otherwise become $\ell$GRBs.
Given that the sGRBs in ellipticals are due to mergers of WDs, 
we can conclude that:
1) the pre-CE/PE impact photon spectrum of $\ell$GRBs is {\it known}, 
2) sGRBs are offset from the centers of their elliptical hosts because 
they are WD-WD mergers in their hosts' GCs (to produce most of their 
MSPs -- Gehrels et al.~2005),
and 3) NS-NS mergers may not make GRBs as we know them, and/or be
%as common as previously thought.\footnote{Without ejected 
%matter these merge within a few ms, far shorter
%than sGRBs \citep{Ei89}.}  This is a 
as common as previously thought,\footnote{Thus
sGRBs may not flag NS-NS mergers, which may last only 
%These may merge in 
a few ms, the same timescale as the 30-Jy DM=375 radio burst
\citep{Lo07}, far shorter than sGRBs (Hansen \& Lyutikov
2001).}
a disappointment to gravitational observatories.
%a few ms, far shorter than sGRBs \citep{Ha01,Lo07}.}  
%This is a 
%a disappointment to the Earth-based gravitational observatories such as 
%LIGO, because 
%sGRBs may not flag NS-NS merger events.

Through their interaction with the overlaying CE and/or PE,
BJs produce the wide variation in GRB/X-ray flash properties 
observed from DD SNe of sufficiently low inclination to the line 
of sight, and the flavors of the 99\% of GRBs due to DD depend 
only on observer inclination, CE and/or PE mass, extent, and 
abundance.
Of the {\it three} different classes of GRBs, $\ell$GRBs, sGRBs, 
and the intermediate time, softest GRBs (iGRBs), as recently 
classified by Horv\'ath et al.~(2006, see also Middleditch 2007), 
most sGRBs occur from DD 
WD-WD merger without CEs or PE, $\ell$GRBs pass through at least 
the PE (necessary for small angle deviations to produce $\sim$100 
s of delay), and usually the CE (which, in addition to the PE, can 
soften the burst), while iGRBs pass through red supergiant (RSG) CEs, 
but little or no PE, possibly the result of a merger of two stars 
with very unequal masses, the possible cause of SN 1993J, which had an
RSG progenitor \citep{Po93}.\footnote{At 1.6\% and 
1.0\% \citep{Tr93} the early polarization of SN 1993J was {\it twice} 
that of the 0.9\% and 0.4\% observed from 87A, 
consistent with even {\it more} axiality than that of 87A.}  
The $\sim$10 s limit for T$_{90}$, and its substantially negative slope 
(tradeoff) with H$_{32}$ for the iGRBs, are consistent with an RSG
CE, but no PE.\footnote{The fluence of {\it both} the 
non-prompt and prompt parts of off-axis $\ell$GRBs are suppressed, the 
first by scattering in the PE, the second by being off axis by the time 
it emerges from the CE, frequently leaving both roughly equally 
attenuated.  This scenario also explains why the two (``precursor'' and 
``delayed'') have similar temporal structure \citep{NP02}.  Negligible 
spectral lag for late ($\sim$10--100 s) emission from ``spikelike'' 
bursts \citep{NB06} can be explained in terms of small angle scattering 
off the PE, without invoking extreme relativistic $\Gamma$'s.} As in 
the case of $\ell$GRBs, the pre-CE impact photon spectrum of iGRBs 
is also known.  

\section{DD in Type Ia/c SNe}
\label{sec:Ia/c}

The list of good reasons against SD for Ia's is long:
(1 \& 2) no SN-ejected or wind-advected H/He \citep{Ma00,Lz02}, (3) 
ubiquitous high velocity features (Mazzali et al.~2005a), 
(4 \& 5) SiII/continuum polarization (CP) both inversely proportional
to luminosity (IPL -- Wang et al.~2006; Middleditch 2006), 
(6) no radio Ia SNe (Panagia et al.~2006), (7) {\it four} Ia's within 
26 years in the merging spiral/elliptical galaxies comprising
NGC 1316 \citep{Im06}, (8) $>$1.2 M$_{\odot}$ of $^{56}$Ni 
in SN 2003fg \citep{Ho06}, (9) cataclysmic variables are explosive
(Scannapieco \& Bildsten 2005), and (10) DD SNe are needed to account 
for the abundance of Zinc \citep{Ko06}.  

{\it No} observation of {\it any} recent SN other than 
SN 1986J\footnote{This SN, in the edge-on spiral galaxy, 
NGC 0891, exceeds the luminosity of the Crab nebula at 
15 GHz by a factor of 200 \citep{Bie04}, and thus
is thought to have occurred because of
iron photodissociation catastrophe 
(FePdC), producing a {\it strongly}
magnetized NS (the origin of magnetic fields in NSs is still poorly
understood, though it is believed that thermonuclear [TN] combustion 
in a massive progenitor to an Fe core is related).}  
and SN 2006gy, including all {\it ever} made of Type Ia SNe, is 
inconsistent with the bipolar geometry of 87A.  Thus, especially in the 
light of SD's serious problems, it seems likely that Ia's are also
DD-initiated SNe, of which some still produce TN yield, but with all 
producing weakly magnetized MSPs.

Further, it seems likely that Ia's and Ic's form a continuous class, 
classified as Ic's when 
viewed from the merger poles, if sufficient matter exists, in excess 
of that lost to CC, to screen the Ia TN products (a rare circumstance 
in ellipticals), because this view will reveal lines of the r-process 
elements characteristic of Ic's.\footnote{As with 87A-like events, it 
would again beg the question of ``What {\it else} they could possibly 
be?,'' 
and ``delayed detonation'' \citep{Kh91}, or 
``gravitationally confined detonation'' \citep{Pl04}, do not produce 
the IPL polarization.  And unless the view {\it is} very near polar, 
this geometry can produce split emission 
line(s) on rare occasions, as was seen in SN 2003jd \citep{Maz05b}, 
and thus again there is no need to invoke exotica, or an entire 
population (III) to account for GRBs (Conselice et al.~2005; M04).}
All this complicates the use of SNe Ia in cosmology, because many Ia/c's in 
actively star-forming galaxies (ASFGs) belong to the continuous class, and 
Ia's in ellipticals (and some in ASFGs) may not produce enough $^{56}$Ni 
to be bolometric \citep{PE01}, lying as much as two whole
magnitudes below the width-luminosity (W-L) relation 
(the faint SNe Ia of Benetti et al.~2005).

A ``missing link'' of Ia's must exist, more luminous than `faint' SNe, 
which fall below the W-L relation by a tenth to a whole magnitude,
may still be largely absent from the local sample, but may {\it not} be 
easily excluded by the TiII $\lambda \lambda$ 4,000-4,500 \AA ~shelf.
There is a more luminous class 
of Ia's, found almost exclusively in ASFGs, that may be attributed to
CE Wolf-Rayet stars (see, e.g., DeMarco
et al.~2003, Howell et al.~2001, and the data in G\'orny \& Tylenda
2000), and a less luminous ``leaner'' class, found in both ellipticals 
and ASFGs \citep{Ha00,Su06,Wa06}, which can be attributed to 
CO-CO WD mergers.  In the DD paradigm, the Ia mass, above the 
1.4 M$_{\odot}$ lost to CC, determines the optical luminosity.  
Since optical afterglows have been found in sGRBs 
with no SNe \citep{GY6,Fyb6,MDV6,Gh6}, DD Ia's can be
very lean indeed.
It is not at all clear if SD {\it ever} happens.

If Ia/c's are indeed the result of the same
process that underlay 10--15 M$_{\odot}$ in 87A, but which 
instead only underlay 0.5 M$_{\odot}$, the outcome will
be even more extreme than the geometry of the SN 1987A remnant.
The PBFs will have higher velocities, and the equatorial/thermonuclear
ball (TNB) will be much brighter, due to the greater concentration 
of $^{56}$Ni.
Thus PBFs form linearly extended structures, whose brightness 
pales in comparison to that of the spheroidal TNBs, which explains 
why Ia continuum and SiII polarization are both IPL \citep{Wa06,M06}, 
and also indicates that part of these lines must originate from the 
sides of the Ia/c PBFs.  

Ia/c PBFs depart and/or thin out quickly because of their high velocities 
and limited masses, potentially exposing a fraction of the TNBs during 
the time interval when $\Delta m_{15}(B)$ is measured.  
Ia/c's with PBFs initially showing r-process lines, because of views closer 
to the poles of the DD merger, are frequently excluded from the local
sample as part of overdiligent attempts to select a ``pure''  sample 
of Ia's.  This selection doesn't work as effectively on the distant 
sample, and the result will be distant SNe which are too faint for the 
redshift of their host galaxies.  Figure 3 of Middleditch (2007) shows 
how this effect could spuriously produce half of $\Omega_\Lambda$=0.7
for a co-inclination (co-i) of 30$^{\circ}$ and a PBF of half angle
of 45$^{\circ}$.  More realistic TNBs which begin as toroids
could produce a big effect even for low co-i's.

\section{Conclusion}
\label{sec:conc}

We have argued that the DD SN 1987A, its beam/jet, 
Mystery Spot, and possible 2.14 ms pulsar remnant, are 
intimately related to as many as 99\% of GRBs, MSPs, and other SNe, 
including all Type Ia SNe, a grave concern for Ia Cosmology.
The time lags, energetics,
and collimation of $\ell$GRBs are consistent with those of 87A's BJ 
and MS, and there is no need to invent exotica, such as collapsars, 
to satisfy them, the expansion of the Universe may not be 
accelerating, and there may be no Dark Energy.  Recent observations
have also cast significant doubt on the existence of dark
matter as well \citep{NP07}.

Given this new, very complex picture of Ia's, any sample, with a very
low dispersion in magnitude, is hardly reassuring.  A rigorous 
treatment of Ia data rules out all cosmologies \citep{Vi05}. 
A straightforward
argument indicates that NS-NS mergers may not make {\it any} GRBs 
as we know them, and/or occur nearly as frequently as previously 
thought.  Models of SNe to date are flawed because neither
the DD process, nor strong magnetic fields have been included,
developments that may still be at least a decade away.
Certainly, no relatively nearby FePdC SN has been well studied, 
SN 1986J having occurred during 1983.  

The DD mechanism ensures that nearly all SNe are born from
a maximally rotating, post-merger WD with a rotation period
near 1.98 s,
thus rapid rotation
can not be invoked as an unusual circumstance, for the
case of SN 2003fg, to justify ``super-Chandrasekhar-mass'' 
WDs (SCMWDs).  The $>$1.2 M$_{\odot}$ of $^{56}$Ni it produced
may only mean that CC underneath mixed TN fuel can initiate very 
efficient combustion/detonation,\footnote{The spectroscopic 
demands of a significant mass of unburned fuel, such as O, being 
invalid because of the invalid paradigm under which such 
estimates were made.} the paltry amounts
of $^{56}$Ni associated with Ib's and at least 90\% of
IIs being the result of dilution of their TN fuel with He
and/or H due to the DD merger process.  Thus SN 2006gy may not 
be a pair-instability SN \citep{Sm06},\footnote{The inner 
layers of all FePdC SNe, possibly {\it many} M$_{\odot}$ of Si, 
Ne, O, and C, have not been diluted with H and/or He by DD, and thus 
may ignite/detonate upon CC, and burn efficiently.  SN modelers 
therefore face the unenviable choice of calculating FePdC SNe, 
which involve strong magnetic fields, or DD SNe, which involve a 
great deal of angular momentum and {\it demand} GRBs as an 
outcome (see $\S$\ref{sec:link}).}
only a massive FePdC SN, which may have produced $\sim$20 M$_{\odot}$ 
of $^{56}$Ni, {\it and} a strongly magnetized NS remnant, a 
prediction which can be tested soon.

Although it would appear that a Universe without collapsars,
pair instability SNe, SCMWDs, and frequent NS-NS mergers which 
make sGRBs, is much less ``exotic'' than previously thought,
SNe themselves are plenty exotic enough, with 1\% producing a 
strongly magnetized NS remnant/pulsar, and the remaining 99\% 
caused by DD, producing $\sim$2 ms pulsars, and BJs which
can incinerate half the planet from a great distance with
little or no warning.  {\it This} is what we will spend a
good deal of the first half of this century figuring out.

\acknowledgments 
I am extremely grateful to CCS-3 for supporting me 
during an interval when I was without funding.  I would 
like to thank Drs.~Aaron Golden, Geoffrey Burbidge, Falk Herwig, 
Peter Nugent, and an anonymous referee for useful suggestions 
which helped me to improve an earlier version of this manuscript.  
I would also like to thank Jerry Jensen for conversations and 
bringing this issue to my attention.  This research was supported
in part by LDRD grant DR2008085 and performed 
under the auspices of the Department of Energy.

\eject

\section{Appendix I:  The Primal Scream Rejection}

Prologue:  The test of a good review is whether the
arguments in it could be sustained in a public forum.
Neither the very long review presented immediately 
below, nor the very short one that follows as Appendix 
II, pass this test:

Dear Jon,

Enclosed please find the EDITED referee's 
report on your submission to the ApJ entitled 
``The SN 1987A Link to Others and Gamma-Ray Bursts'' 
(MS\# 72836).

Please don't let the occasional harsh words in the 
report distract from the point that the referee 
identifies many critical flaws in the manuscript. 
It seems to me that withdrawal would be the best 
option. If you feel that the referee does not 
understand your work well enough, you could request 
a new and independent referee. I would support that, 
BUT if the 2nd report is negative as well I would 
have no choice but to reject the paper.

Please let me know asap whether or not you agree 
with my suggestion to withdraw the submission.

If you have any questions, feel free to contact me.

I am sorry the news did not turn out to be more favorable.
With best regards, Dieter

*************** Referee Report (with reply):

Whoa!  A 2nd report?!  That means I can get another gem 
like {\bf this} one?!

Well duh, Dieter.  This person is obviously not objective 
about a paper that turns much of that reviewer's research 
into vapor.  ``This is not exactly a review, more of a primal 
scream.'' -- a colleague.  I am surprised that ApJL even 
considers this valid.

Many of the reviewer's objections are seriously out 
of date (Thielemann 1990?), and the rest have valid rebuttals.  
These issues are addressed in full detail below (the specific
small changes made to the manuscript are not detailed herein.

This is my answer to the points made by this referee:

A common thread runs through the criticism of \#72836.  That thread 
is that calculations, or the lack thereof, trumps straightforward 
interpretations of observations.  It is far too early, and the 
potential calculations far too complicated, to use their absence as 
criticism of observational interpretation.  An example is the 
collapsar computations of MacFadyen and Woosley, which allow 
astronomers to continue on and claim that they are actually doing 
real science.  I have a lot of respect for those who chose to hack 
away at such a difficult endeavor, but it's a perversion of the 
science to use them to filter out unpopular interpretations of 
observations.  Calculations are basically theory injected into 
computers, and will NEVER fail to err without rigorous observational 
constraint, and I'm not talking about just a single split emission 
line from one SN.

It is indeed unfortunate, possibly even tragic, that an entire 
generation of young astronomers were misled by their mentors, 
because those mentors rushed to ignore the implications of valid 
observations of SN 1987A, even those which had been reproduced by 
more than one competent group of observers.  Now a large segment 
of this generation is IN THE WAY, using up large telescope time in 
the fruitless search for Dark Matter and Dark Energy, neither of 
which will ever be found to exist (see Nelson \& Petrillo 2007 on 
the absence of Dark Matter).  ``What they should be doing is studying 
the nearby SNe, on the smaller telescopes, but they don't want to do 
that.'' -- a colleague.  In the meantime, SN 1987A, the one nearby SN 
in nearly 400 years, has not been observed in high time resolution by 
anyone for over 11 years, and that's criminal.  The equatorial ring 
will be a factor of 10 brighter in 10 years, and yet another factor of 
10 in the next 10.  Time is running out.  Far worse still, is that by 
holding out hope for new physics, this frenzy may be distracting 
political leaders from taking steps to mitigate global warming, and in 
that sense it is irresponsible for them to remain in their state of denial.

	{\it In this paper, John Middleditch proposes, using SN 1987A as
	his prime example, that essentially all supernova-like events
	are caused by the double-degenerate merger of two white
	dwarfs, presumably CO white dwarfs, leading to the formation
	of a rapidly spinning millisecond (ms) pulsar. The author
	suggests this as a unifying scheme that can explain all these
	events, which he finds intellectually attractive, although
	this is an aguable [sic] point, considering the diversity of
	observed explosions.

	While the paper makes a couple of interesting points that
	could stimulate discussions in the field, some of the main
	claims are either wrong or unproven and can only be put
	forward by ignoring the wealth of detailed scientific
	literature (see below for more details). Because of this, I do
	not think that this paper can be published. It would be
	acceptable to have a speculative paper that stimulates ideas
	in a new area of research, but ignoring well established facts
	and a whole section of the relevant literature is
	scientifically unacceptable.}

Au contraire, my suggestions are quickly becoming the established 
norm among most of those working in the many subfields (see below).

	{\it The main idea of the author is that a double-degenerate merger
	can produce a diversity of observable supernovae depending on
	whether it is surrounded by a common envelope or depending on
	its viewing angle. While it is plausible that this leads to
	different observational events, the link of this postulate to
	observed supernova types is less than convincing. It is
	generally believed (for good scientific reasons!) that there
	are at least two different explosion mechanisms, core collapse
	and thermonuclear explosions (leaving GRBs aside for the
	moment).}

Toward the end of the Wednesday afternoon session of the SN 1987A, 
20 Years After and GRB conference in Aspen on Type Ia SNe, the 
question was asked: ``Is there any way of avoiding double-degenerate 
for these [Type Ia SNe]?''  Someone ventured an answer, but Nino 
reminded him that his suggestion had already been discredited.  
There was no other reply.

Bob Kirshner was there.

Craig Wheeler was there,

Alex Filippenko was there.

Tom Janka was there.

So the current thinking on Ia's is that they ARE indeed DD, which means 
that they are core-collapse objects producing NSs, which likely will 
indeed be weakly magnetized and rapidly spinning.

What does THAT say about calculations of ``gravitationally 
confined detonation''?

What does that say about calculations of ``delayed detonation''.

For that matter, what does it say about ``collapsars'' being real?
	
And if Ia's are DD, why not other SNe of progenitors of modest 
mass, which share at least early polarization in common?

	{\it The author seems to invoke only core collapse even to
	explain SNe Ia that on average eject 0.6 Msun of Ni and
	sometimes substantially more. How is this possible in this
	scenario? The author briefly addresses this issue in the third
	but last paragraph, but that discussion is little more than
	uninformed gobbledegook and reveals an astonishing lack of
	understanding of basic supernova physics. All of a sudden he
	refers to ``efficient combustion/detonation'' to produce core
	collapse, i.e.  a ms pulsar, and 1.2 Msun of Ni? These numbers
	just do not add up. He now introduces the concept that
	different ways of mixing ``TN fuel'' produce different types of
	events. None of this discussion refers to any proper
	simulation of double-degenerate mergers or makes any
	suggestion for the cause of the differences from event to
	event. In this context, it should be noted that there has been
	substantial progress in understanding thermonuclear explosions
	from first principles that are generally believed to produce
	SNe Ia. While there are still some detailed arguments
	concerning, in particular, the transition to a detonation, the
	basic paradigm is very sound and cannot just be ignored.}

The single degenerate paradigm for Ia's is lying on the floor, shattered 
into {\bf pieces} (see above), so this criticism is irrelevant.

Again, also, the science is being perverted.  A super-Chandrasekhar mass 
WD is a fundamental violation of known physics, and its existence should 
require extraordinary evidence, not just so much blather about unburned 
mass determined from some lines, especially if these were interpreted 
in the context of a {\bf wrong} paradigm.

The inference of the spectroscopy as regards to amount of unburned material 
has {\bf never} been redone in context of DD, and the resulting bipolar SNe.  
The high luminosity of the thermonuclear ball is no guarantee because the 
polar cones/jets can shade/expose it.

	{\it There is clearly a lot of confusion, even in the author's mind,
	what he means by a double-degenerate merger, but let me
	now address the various sections more systematically.}

The folks at Aspen had no problems with it.

	{\it SN 1987A:

	The author uses SN 1987A as his prime example for a
	double-degenerate scenario. He points out correctly that this
	was an unusual event, specifically referring to the bipolar
	nebula and the mystery spot, both of which could be indicative
	of rapid rotation. Indeed, he points out that the probably
	most promising model for the progenitor invokes the merger of
	two stars, though not as he claims of two electron-degenerate
	cores but of a red supergiant with a normal-type star. A
	double-degenerate merger inside a common envelope, as he
	proposes, could possibly also explain some of these supernova
	features, but it cannot explain many others. First, the merger
	occurred 20000 years before the explosion (based on the
	dynamical age of the nebula).  Why would the star look like a
	blue supergiant for 20000 years?  ... }
	
Easy -- too much angular momentum.  A blue straggler on 
steroids.  Does this reviewer seriously believe that SN 1987A 
had an Fe core?
	
	{\it Even if the core collapse
	event could be delayed by 20000 years (this may well be
	possible), the merged cores surrounded by a large envelope
	would almost certainly have the appearance of a red
	supergiant, just like any 10 Msun star with a compact core of
	about 2-3 Msun, not a blue supergiant (if the author is
	convinced otherwise, he would have to demonstrate this by a
	reasonable calculation).}

Again, too much angular momentum.  ApJL is a journal where one 
can make reasonable suggestions, without having to take a year each 
to run calculations (which have been discredited anyway -- see above) 
on every little detail.  In addition, are current calculations even 
capable of resolving this question?  I  doubt it.  It is not even 
clear what the criticism is about.  The reviewer is just trying to 
stall this paper, in case he can't kill it, which is embarrassing 
because he/she and so many others were so clueless for such a long 
time.  Ia's have been found to be DD at Aspen, and at Santa Barbara, 
they were apologetic about it, and the whole house of cards is still 
collapsing as I write.  I don't OWE them waiting until 20 minutes after 
THEY decide the paradigm has shifted, before {\bf I} can write about it.

	{\it I have used these particular values,
	trying to imagine how such a scenario could work, taking a
	positive constructive view, but this reflects another generic
	problem with the paper, namely that it is lacking enough
	details to allow a proper evaluation.}

Again, it is easy to suggest that detailed calculations be made every time 
a suggestion is offered, with the full knowledge that ApJL is not the place 
where there is room to do this.  And again, it's a stalling tactic.

	{\it The second, even more
	severe problem is that we know from the analysis of the
	supernova ejecta that the core of the star that exploded had a
	H-deficient core of at least 5 Msun (see, e.g. the work by
	Thielemann 1990, ApJ, 349, 222, but also many [!] other people
	like Arnett, Woosley).  This is not compatible with a
	double-degenerate merger that can produce at most 1.5 Msun of
	non-H ejecta (assuming that 1.5 Msun went into a neutron
	star).}

A 1990 paper is pretty much out of date.  Woosley, Burrows and others 
at the Aspen conference made no attempt to eliminate, or discredit DD 
as a possible hypothesis for SN 1987A.  Tom Janka paid tribute to Philipp's 
and Morris's work on the SN 1987A rings, and its implications supporting 
binary merger (and he still went on to use an Fe core, because no one's 
{\bf ready} to calculate DD).

	{\it Again, if the author had good scientific reasons to
	challenge the other studies, it would be up to him to
	demonstrate why these detailed studies are wrong or at least,
	at the very minimum, show that his model can produce the
	basic, observed characteristics of the SN 1987A ejecta.
	Anything else only qualifies as a fantasy product, not
	science.}

The reviewer is trying to draw the line here about DD for 87A, but 
with most or all Ia's DD, it just doesn't wash.  In Iabc's the 
dominant mechanisms will be evolution- or collision-induced mergers.  
In IIs, evolution-induced merger will dominate. So what? With the 
rings, the bipolar explosion, the mixing of the elements, the blue SG, 
the Mystery Spot and its coincidence with the early light curve, and, 
YES, the 2.14 pulsar (see below).  If SN 1987A wasn't a DD SN, I don't 
know any more clues about it a SN can possibly have.  A full treatment 
of SN 1987A ejecta under the DD paradigm is a huge task, and will take 
a few years, and is homework for the modelers.

	{\it GRBs:

	The author also tries to link SN 1987A to GRBs more generally.
	He points out, correctly, that the DD process must dominate by
	a large factor over NS-NS mergers in early-type galaxies and
	then continues that this ``leads to the inescapable conclusion
	that the DD process produces sGRBs''. I am sorry, but this is a
	simple non-sequitur, since it would first need to be
	demonstrated that DD mergers can produce a GRB in the first
	place. Again the author ignores detailed work on
	double-degenerate mergers (e.g. by Rosswog, Janka and others)
	and how this can lead to a truly relativistic
	event. Nevertheless, in the area of GRBs the uncertainties are
	large enough that a DD merger can probably not be ruled out
	(the author may want to look at the work by Todd Thompson and
	collaborators since that could potentially support some of his
	suggestions), but the logic as presented is not really
	tenable.}

THIS is the argument that drove Rejean Dupuis out of LIGO and into 
a banking career!  And also where the screaming gets loudest (see the 
appendix to astro-ph/0608386).  See above for 87A being due to a DD 
merger -- it's credible enough, certainly by the standards of a 
subfield where collapsars get accepted where there is no compelling 
evidence for them whatsoever.  Events (Lorimer et al. 2007) have 
proven me correct on this account, as NS-NS mergers have a
different signature than GRBs.

This reviewer just says he doesn't like the logic, but really never 
says why.  Much of this is just a version of name-calling.  Tom Janka 
was in the audience when I gave this talk that Tuesday morning at Aspen.  
He didn't challenge it.  We were both around all the rest of that long 
day, looking at each other.  Tom never engaged me on these issues.  
The logic of Horvath et al. 2006 applies BOTH ways -- they did not see 
the need for any other subclass of sGRBs on the low T90 side.  
So $>$90\% of them are WD-WD mergers, or DD.  Who really believes there 
could be that many NS-NS mergers?  Aside from that, the 30-Jy, $\sim$5 ms, 
DM=375 radio burst, http://www.sciencemag.org/cgi/content/full/318/5851/777,
is an obvious candidate for a NS-NS merger, the $\sim$100/day/Gpc$^3$ rate 
being consistent with Kalogera's estimate of ~3/day/Gpc$63$ when 
inspiraling DTN NS-NS pairs are counted.

	{\it SNe Ia:

	The logic gets even more mangled in the next section, since
	all of a sudden the same DD mergers are also invoked to
	account for SNe Ia and normal SNe Ic. The author does not
	really explain how DD mergers produce this diversity, except
	to vaguely refer to BJs, PEs, MSs and other abbreviations the
	author introduces, which mainly serve to obfuscate the logic
	of the paper. The author postulates further that the
	difference between a SN Ia and Ic is one of viewing
	angle. While at early times this might be a possible
	interpretation, it is not at late times; in the late nebular
	phase the ejecta are transparent, and it is straightforward to
	measure the total amount of Ni produced in the supernova from
	the nebular Fe lines, irrespective of viewing angle. This has
	been done extensively for nearby supernovae (e.g. the work by
	Meikle, also Mazzali et al.), demonstrating that the
	differences of these supernova types cannot just be due to
	viewing angle. This example illustrates quite blatantly that
	the author does not really know much about supernova research.}

At late times the spectra of Ia's and Ic's are nearly identical.  Again, 
interpretations assuming spherical geometry for bipolar explosions may 
be seriously in error.  Mazzali may have made the claim that Ia's can't 
be Ic's when viewed from the poles, but he's also one of the number who 
stated about Ic's in abstracts that they ``must be massive stars,'' a
claim that has not held up, and one which Craig Wheeler has also cautioned 
against.  Mazzali should be praised for his observations, and {\bf beaten} 
for his abstracts.  Aside from that, Ia's are DD, NOT thermonuclear 
disruption.

	{\it Are all NSs formed as ms pulsars?

	The author's model implies that most neutron stars are born as
	millisecond pulsars. Indeed, I remember that in an earlier
	version of this paper (on astro-ph) the author claimed that
	this was the case. Even though this section has been removed
	(for good reasons), this implication does not appear
	consistent with what we know about radio pulsars. Again the
	author would have to ignore a whole wealth of literature on
	this topic, in particular from the recent Parkes multi-beam
	survey. In this context, the author claims that the standard
	recycling scenario for ms pulsars does not work.  Here he
	ignores recent discoveries of ms pulsars in X-ray binaries,
	i.e. in the process of being recycled (see in particular the
	work by Bhattacharya), which has quite impressively confirmed
	the basic recycling paradigm. The only thing the author could
	challenge is the recycling + evaporation scenario for the
	production of *single* ms pulsars for which a DD collapse
	provides a respectable alternative.}

The point X-ray source in Cas A is radio quiet.  So much for radio 
pulsars and the logic of the lamppost.  TeraGauss NSs appear to be 
a rare minority, and this is supported by the lack of hot centers in 
recent nearby SNe, where SN 1986J is so far the only known exception 
(and we will know soon enough for the case of SN 2006gy).

The recycling scenario has been on the ropes since I found the 1st
MSP in a gobular cluster two decades ago, and X-ray MSPs don't save 
it.  I asked Fred Lamb whether we knew these pulsars were recycled 
or born fast, and his answer was the we don't know, and I don't 
think his mind has changed since.  There's no way of telling how an 
MSP was formed, whether born that way, most likely in a 
binary-binary merger-induced core-collapse, or recycled from a 
companion acquired after it has died as regular TeraGauss pulsar.  
However this last possibility requires field decay, or at least 
effective dipole reduction, because its magnetic poles have migrated 
to opposite sides of one of the magnetic poles ala Chen and Ruderman 
(1993), but it's not clear whether such a geometry can be as easily 
recycled as a truly weak magnetic field.  The binary-binary 
collision-produced MSP can inherit a companion from the process, no 
need to go looking, and field decay is not required.  And THESE can 
be recycled from such companions, and there's no way to tell the 
difference.

This paragraph also ignores the implications of two recycled pulsars 
in binaries in Ter 5, with the fastest at 10 ms and change, both of which 
weighed in at 1.7 solar (a similar situation holds for M5).  Also from 
Chen, Middleditch, and Ruderman 1993, the pulsars in the core-collapsed 
GCs and non core-collapsed GCs also indicate that recycling can't get 
TeraGauss pulsars into the true ms range.  With injection from 2 ms DD 
PSRs, which already have weak magnetic fields, recycling can reduce 
their periods below 2 ms, a validation for Ghosh and Lamb 1979, without
requiring field decay.

	{\it Logic of Presentation:

	I have already indicated some serious problems of logic
	in the paper, but there are many more instances where
	the author makes claims without any substantiation of the
	claim (or valid reference). In several cases, the author
	gives references, but seriously mis-quotes the papers he
	is referring to. This in itself is scientifically
	unacceptable.

	Here are a few examples:

	o Introduction: claim that the beaming factor of GRBs is 10$^5$.
	The only reference given is to a paper by Meszaros who does
	not claim this. Reasonable estimates are 100 to 10$^3$.
	I believe there is one paper by Don Lamb suggesting a much
	larger beaming factor but that claim has not survived further
	scrutiny, as even Don Lamb has implicitly admitted in later
	papers.}

The behavior of the light curve of 87A around days 7-10 indicates that 
the beam spot and particle jet spot on the polar ejecta must be smaller 
than 1 lt-day.  The distance is $\sim$20 lt-days, so the beaming factor 
for 87A was higher than 10,000.  The paper was so modified, and there 
is no need to solicit anyone's opinion.  Anyway, one suspects that the 
motivation to revise the beaming factor downward, at least for lGRBs, 
is just to substantiate the claim that they result from exotic, hence 
rare (or vice-versa), events, thus keeping the hyperbole flowing.  

	{\it o Footnote 1: claim that the ``discovery'' of a 2.14 ms pulsar
	in the SN 87A remnant ``is no longer controversial''. I beg
	to differ. It has not been seen by other groups who should
	have been able to see it since they were looking at similar
	times.}

The story of high time resolution observations of 87A made by others is a is
sorry tale of incompetence and inadequate effort.

The object was found in data from many telescopes and observatories.  Sure, I 
did the first pass analyses, but reputable collaborators have also verified the 
signals, and the data have been offered to {\it anyone} who requests it.  The 
probabilities in Middleditch et al. 2000 are generous enough.  They are not off 
EIGHT orders of magnitude.  The Tassies didn't hallucinate their data.  Who is 
this reviewer that he thinks he can ignore OUR publised observations, while 
claiming that I can't ignore flawed and/or inadequate/non-existent observations 
of others?

We've even had a night in common with another group, with an agreement to share 
the data (I have a {\bf slide} of the guy with Jerry Kristian on the afternoon 
of 1992, Nov. 6, in the Las Campanas 2.5-m control room).  The promised data was 
never delivered, even though we did ask for it.  (WHY?  Written over inside the 
laptop?  Lost?  Absolute verification of the signal too damaging to astronomers?  
Ergo decades of work down the drain?  Likely written over )

The guy said ``I don't see much.'' Kristian said ``That's not real helpful.''

By those standards, we didn't see MUCH, but there was {\bf something} there, 
and a common observation night, even with a less restrictive filter would have 
told us a LOT.  We used a Wratten 87A (basically the I band, 800-900 nm).  
The guy used a GG495, basically a 500 nm longpass.  There is a factor of 10 
difference in 
count rate on 87A between these two on the same telescope.  If the signal 
were present for the entire GG495 band, he would have seen 20 times the power 
that we saw.  If it were restricted to the Wratten 87, we would have seen a 
factor of 5 times more power than he saw.  We never could convince ourselves 
that including the 500-800 nm band {\bf ever} did anyone any good.  But isn't 
that what simultaneous observations CAN do for us?  What an unforgivable waste!

I also pointed the HST/HSP collaboration to candidate frequencies for which we 
found a signal in their data on June 2, 1992, and March 6, 1993.  For whatever 
reason they did not respond then, wrote a paper claiming an upper limit of 27th 
magnitude, which Kristian and I refereed.  We informed them that it was really 
22nd (100 times brighter).  The HSP count rate on any object of known magnitude 
will verify that the instrumental throughput is 1\% at best, and from this, limits 
can be set from the total number of counts in ANY observation.  We told ApJ 
that we'd like to see the paper again before it got published.  Next time we saw 
the paper it {\bf was} published with a limit of 24.5 (still exaggerated 10 times too 
dim).  A representative of the collaboration showed up at the SN 1987A -- 10 Years 
After conference in La Serena, Chile, and tried to argue for this limit, at least 
until he showed a power spectrum of his calibration object.  When I informed him that 
the object had 10 times less background than SN 1987A, and was also integrated over a 
50\% longer time interval, he could only leave the stage, muttering.  Also, a total 
of 160 minutes of observations of SN 1987A in a couple or YEARS?  It's like 
they planned to FAIL!

Manchester and Peterson published on not seeing a signal in Dec. of `94.  I looked at 
their data, they {\bf really} didn't see anything.  But they spent only part of the 
two nights on 87A.  In fact, at the Aspen SN 1987 \& GRBs Conference during Feb. 19-23, 
over a dozen YEARS after this observation (with none in between as far and I know), 
Manchester made a whole contributed talk on the basis of this observation, plus the 
published times 10 exaggerated faint limit (24.5) from the HSP.  As I had done a 
decade earlier in La Serena, I had to correct the exaggeration on the spot.  Mark 
Phillips remarked to me afterward, ``I thought he was there!'' So did I!  The kindest 
thing which can be said is that he went off somewhere for that part (he was 
certainly there prior to that).  Also clearly, Manchester had never {\bf looked} at HSP 
data, much like a lot of other people who THINK they know what the answer is about SN 
1987A.  This was not even corrected in Manchester's proceedings contribution to the 
Aspen Conference.  It quotes the HSP limit as ``$\sim$ 24'' -- 24 to 22 is quite a 
stretch even for a '$\sim$'!  He also quotes his limit on 87A from 4 100-minute segments 
during his and Peterson's 2 nights on the AAT in 1994, early Dec., the last time he ever 
observed 87A, at 24.6.  We observed 87A with the CTIO 4-m for 18.6 hours in 1993, late 
Dec., and achieved a limit of 24.0, detecting the 2.14 ms signal on all three nights at 
24.77(0.2), 24.44(0.3), and 24.78(0.2) in the V, R, and I combined bands (using a gold 
secondary) about 2/3rds of the count rate of an aluminized secondary.  So his 24.6 limit 
is likely exaggerated by at least a magnitude.

Aside from that, it's not like there {\it aren't} 10 solar masses of starguts moving around 
(also no guarantee that the pulsar remnant isn't precessing and potentially changing its 
beaming).  As far as I know, they also had no observations during the interval from Feb. 
of `92 through Sep. of `93 (they tried on September 15, 1993, but were clouded out -- 
signals were seen from Tasmania on the 12th and 24th), when we were detecting the signal 
most consistently.  Remember, HST was still nearsighted during that interval.

So THAT was our competition.

	{\it o Section 2. ``PBF - the prime suspect for the r-process'';
	not necessarily wrong, but a statement without reference
	or explanation.}

OK, fine.  `` ... (PBF -- a needed candidate for the r-process, e.g., Arnould et al. 2007) ... '' 
Arnould, Goriely, \& Takahashi 2007: ``After some fifty years of research on this subject, the 
identification of a fully convincing r-process astrophysical site remains an elusive dream.''

	{\it o Footnote 8: Claim that NS-NS mergers occur within a few ms.
	A reference is given, but this one is outdated. Recent
	detailed simulations by Ruffert, Janka, Rosswog et al. have
	shown that this is not the case.}

Hard numbers are remarkably absent from the abstract of their latest paper 
(III) on NS-NS mergers.  Tom Janka was in the audience at Aspen when I gave 
my talk, including the bit about sGRBs being predominantly DD events.  He 
did not comment then, and has not commented since.

	{\it o Footnote 12: ``What else could they be?'' plus ``there is
	not need to invoke exotica...'' This sounds like desperation
	rather than well-founded scientific argumentation. Indeed,
	what is exotic to one person may not be exotic to another
	person.}

Not desperation, but Occam's Razor, a principle which, unfortunately, has been 
absent from much of the garbage that astronomers are promulgating in this subfield, 
PLUS the fundamental principles of astronomy itself --  sources, parents, 
offspring, etc.  Occam's Razor cuts both ways. What DO Ia's look like when viewed 
from their merger poles?  If the MS of SN 1987A wasn't related to GRBs, then what 
WAS it?  It HAS to look pretty impressive when viewed from either pole, likely 
visible at cosmological distances.  Where are THESE events in such samples?

	{\it In summary, I do not believe that the author presented a
	consistent and coherent case that all supernovae are
	related to DD mergers. Considering how much he has ignored
	the published literature, I do not think this paper should
	be published in a respectable journal.

	Could the paper be modified to make it publishable? In
	principle, the author could revise the paper by removing some
	of its logical inconsistencies and addressing the relevant
	published literature. However, it would not be enough to just
	point out that there are numerous unresolved uncertainties in
	these models (which is definitely true). But since it is this
	author who it going out on the limb, it would be up to him to
	demonstrate by performing reasonable model calculations that
	DD mergers can account for the phenomena he is invoking them
	for. These would not have to be state-of-the-art
	multi-dimensional hydrodynamical simulations, but at least
	have to contain enough reasonable physics to support the
	author's claims. Without these, the paper falls more in the
	category of a phantasy novel than a piece of respectable
	science. Considering that I judge many of the author's key
	claims to be false, I doubt that he would be able to satisfy
	these requirements.}

It is not I who is living in a ``phantasy'' world, but this reviewer 
and his/her brethren.  Collapsars, SCMWDs, Dark Energy, pair instability 
SNe, and likely even Dark Matter and will all be found to be garbage 
(there is a still a chance that collapsars have something to do with 
black hole formation, but because DD can make, and extreme energetics are 
no longer required for, GRBs, collapsars are no longer needed to 
make GRBs), and SN 1987A provides the leverage through which this will 
be accomplished.  If not now, then when?  If not by me, then by whom?  
I was right about the GRBs, I am being found right about MSPs,
and I will be found to be right about SNe, and the pulsars they
leave.  And all this simply because I am rational, when many are 
not, and have a sense of what is garbage, and what is not.

Do these people care what the truth is, or have they abandoned that 
concept so that they can make their lives easier by doing things they 
are used to doing, instead of those that really need doing, but are much 
more challenging?  At this level, the effects of academia mixed with a 
defunct grant system is {\bf preventing} progress from being made in astronomy.  
Again ``What they should be doing is studying the nearby SNe, on the 
smaller telescopes, but they don't want to do that.'' -- a colleague.  
The revelations of this work show that much of astronomy has to
be rebuilt from the ground up, and I can think of no better use
for today's graduate students.
These folks have monopolized the big telescopes for the last decade or so, 
and think that hard work alone is enough to merit a continuance of this 
state of affairs, NO MATTER how the science breaks.

Like the DD issue for Ia's, their paradigm(s) have/are crumbled/crumbling 
beneath their feet.  The leaders of this crowd are getting around to 
admitting it -- they wouldn't have survived the last few conferences 
if they hadn't.  However, the message apparently has not filtered down to 
their followers (and this referee is among them), who act as if their 
audacious, and scientifically unsound assertions haven't already been 
seriously challenged.  They wouldn't be so afraid of one dissenting 
paper if their own case weren't already toppling around their ears like 
a house of cards.

When {\bf I} knew I had a spurious result, I retracted it.  After Dark Energy 
and a lot of other stuff is found to be garbage, these folks will likely 
just slink back to the halls of academe.  Tom Siegfried's take on the Santa 
Barbara SN meeting:

http://www.sciencemag.org/cgi/content/full/316/5822/194, 
is likely all that will happen.  Support to do this bad astronomy amounts to 
WPA for astronomers.  After all of this hullabaloo, astronomers will be lucky 
if anyone is still willing to give them funding.  If they take yet another decade 
to get around to admitting their problems, then likely no one will give ANY of us 
funding EVER again.

\eject
\section{Appendix II:  The One Paragraph Diss Rejection}

Dear John,

I have received a report from the referee on your revised 
ApJL paper cited above. A copy of the report is appended below.

The referee finds significant problems with your paper and 
recommends against publication. In view of the referee's 
assessment of your paper, and the negative report of the 
previous version of the paper, we will not be able to accept 
this paper for publication in the ApJ. 

I am sorry that the revisions did not lead to an acceptable 
paper. The referee is a very experienced, and also a very 
objective person (willing to go some distance on topics that 
are ``out of the box'' or ``non-mainstream''), but the assesment 
was still negative - with very strong words regarding the 
impossibility of a further revision leading to an acceptable 
paper.  
   
I hope you can find another appropriate venue to promote (as 
in publish) the ideas presented in your manuscript. 

With best wishes: Dieter

--------------------------------------------------

It is easy to sit there and contend that the assertions are 
unsupported, the logic is vague and elusive, and that there 
is little evidence.  It is easy to sit there and claim the 
arguments are too tenuous, knowing full well that this is
the most developed possible set which still fits within the 
space alloted for an ApJ letter.  

In fact the review itself is what is vague, elusive, and 
presents unsupported arguments, a classic case of the pot
calling the kettle black.  And oh my!  How tired one 
must get when encountering the dread `SCMWD' for the 2nd time!  
(SCM means something else to me.)  
We don't know how this beam/jet from Hell was formed, we
only have that picture of 87A, its early light curve, and
data on the Mystery Spot.  Most of us don't know yet how 
pulsars shine.  So what?  That comes later, in a bigger 
paper.  Requesting an explicit mechanism is just another 
way of stalling.

As vague and indefinite as it is, it correctly predicted 
that NS-NS mergers do not make GRBs (see above and Lorimer et 
al.~2007), the bimodality of the masses of MSPs in globular 
clusters (Freire et al. arXiv:0712.3826), the offsets of sGRBs
from the centers of their host elliptical galaxies, the details 
of Ia's, including their two faint subclasses,
high velocity features, inverse relation between polarization 
and luminosity, and also makes a explicit prediction as to the 
cause and outcome of SN 2006gy.   This pile of vague, unsupported 
objections, disguised (poorly) as a review, is a symptom of what 
this branch of astronomy has become -- a disengenuous exercise
perpetrated on the American taxpayers so that astronomers can 
pretend their paradigms haven't crumbled, and BS until the end of 
the Universe, allowing tiny little dollops of progress only when 
everyone has covered their behind about having been so utterly 
clueless about SNe, GRBs, and MSPs.  They can't argue these 
points in public, outside of the cloak of anonymity provided 
by the journal, as there really is no rebuttal to them.  

{\it I have read this paper three times with good will and a generous 
approach to lively scientific discussion. I reluctantly conclude 
that this paper does not meet the standards of the Astrophysical 
Journal. Although it refers to many interesting astronomical 
phenomena, the conclusions do not follow from the evidence and 
there is precious little evidence. If this paper contained a cogent 
and quantitative physical discussion of the way in which the 
observed phenomena in SN 1987A shown in Figure 1 are plausibly the 
result of the mass loss followed by beamed ejections from that 
object, it might possibly be suitable for the ApJ, but the present 
discussion is a series of qualitative unsupported assertions, 
followed by unjustified leaps to unrelated phenomena.  The paper is 
almost impossible to read, due to a propensity to use novel 
abbreviations (SCMWD) for phrases repeated only a few times. There 
is a nugget of a scientific idea here, trying to unite a wide 
variety of phenomena with the notion of double degenerate mergers. 
But the evidence presented is so fragmentary and allusive that it 
does not constitute a scientific case for any of the proposals made 
here. It would be a mistake to publish this paper in the 
Astrophysical Journal. It would be a mistake to impose further on 
the editorial processes of the Journal and the goodwill of the 
scientific community by offering a revised version of this paper. 
I will not serve again as a referee for this paper.}

\begin{figure}
\vskip 7 in
\includegraphics{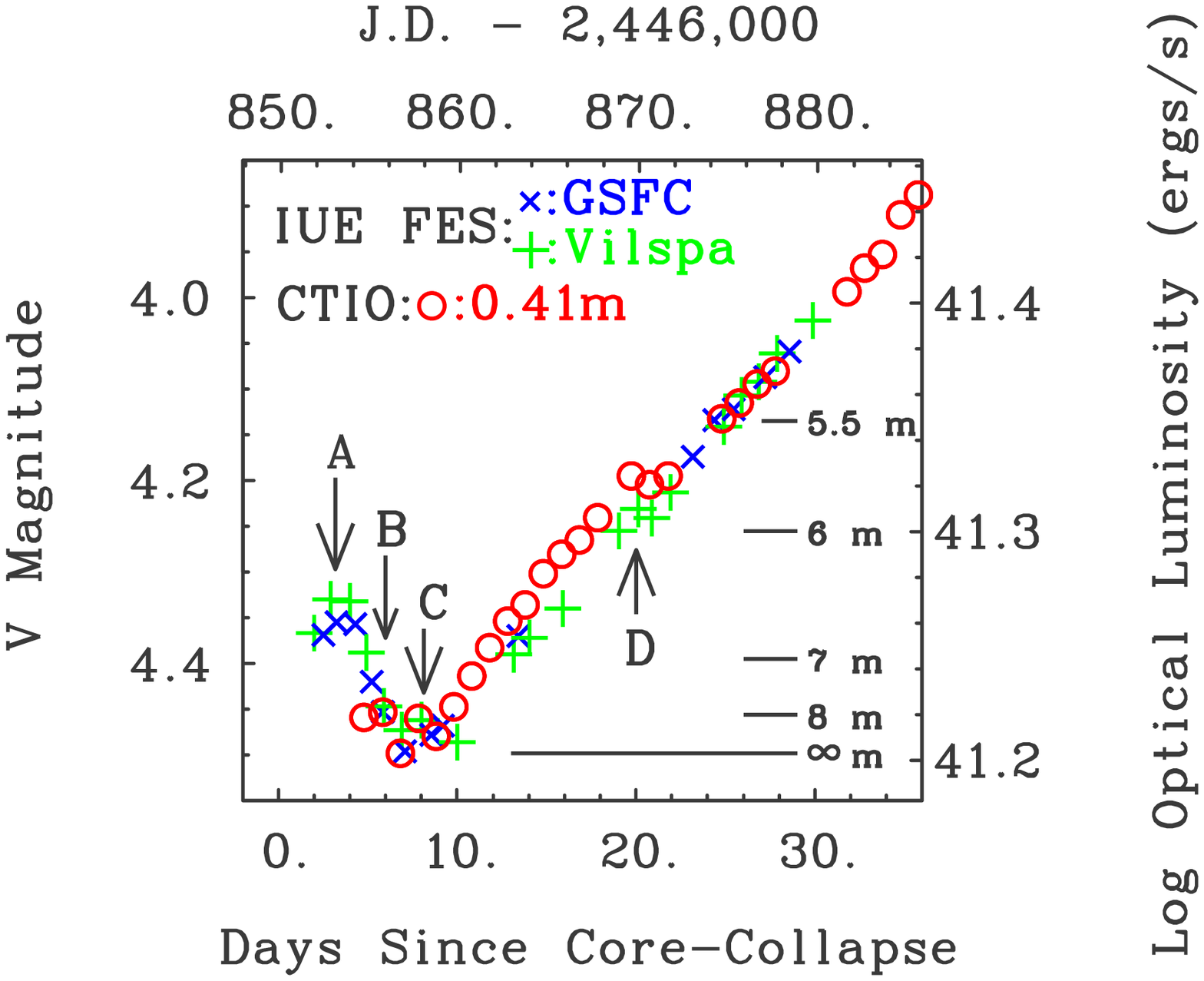}
%\epsscale{0.02}
%\plotone{f1.eps}
%\caption{The very early luminosity history of SN 1987A as observed 
\figcaption{The very early luminosity history of SN 1987A as observed 
with the Fine Error Sensor of IUE and the 0.41-m at CTIO.  Data 
points taken at Goddard Space Flight Center by Sonneborn \& Kirshner, 
the Villafranca Station in Madrid, Spain, are marked (see 
$\S$\ref{sec:early}).  
%Various stages of beam/jet breakout and 
%interaction with polar ejecta are labeled 
%\figcaption{The very early luminosity history of SN 1987A as observed 
%with the Fine Error Sensor of IUE.  Data points taken at Goddard Space 
%Flight Center by Sonneborn \& Kirshner, and the Villafranca Station in 
%Madrid, Spain, are marked.  Various stages of beam/jet breakout and 
%interaction with polar ejecta are labeled see $\S$\ref{sec:early}.
%The fit to the six points 
%from day 854.5 to 857 is a 
%parabola, consistent with optically thin thermal radiative cooling.
%The decrement near day 20 is actually preceded by a {\it spike}
%with strange colors (B, R, \& I, but little U or V -- see the previous
%viewgraph -- a reverse shock? pulsar?).
     }
\label{fig:FES}
\end{figure}

\end{document}